\documentclass{iau}

\usepackage{amsmath}
\usepackage{graphicx}
\usepackage{multirow}

\begin{document}

\lefttitle{X. Fang et al.}
\righttitle{GTC Spectroscopic Surveys of Planetary Nebulae}

\journaltitle{Planetary Nebulae: a Universal Toolbox in the Era of Precision Astrophysics}
\jnlDoiYr{2023}
\doival{10.1017/xxxxx}
\volno{384}

\aopheadtitle{Proceedings of IAU Symposium}
\editors{O. De Marco, A. Zijlstra, R. Szczerba, eds.}
 
\title{GTC Spectroscopic Surveys of Planetary Nebulae \\
in the Milky Way and M31}

\author{Xuan Fang$^{1,2}$, Haomiao Huang$^{1,3}$, 
Mart\'{i}n A.\ Guerrero$^{4}$, 
Letizia Stanghellini$^{5}$, 
Rub\'{e}n Garc\'{i}a-Benito$^{4}$, 
Ting-Hui Lee$^{6}$, 
Yong Zhang$^{7,2}$}
\affiliation{
$^{1}$National Astronomical Observatories, Chinese Academy of Sciences (NAOC), Beijing 100101, China; \email{fangx@nao.cas.cn} \\
$^{2}$Laboratory for Space Research, Faculty of Science, The University of Hong Kong, Hong Kong, China \\
$^{3}$School of Astronomy and Space Science, University of Chinese Academy of Sciences, Beijing 100049, China \\
$^{4}$Instituto de Astrof\'\i sica de Andaluc\'\i a, CSIC, Glorieta de la Astronom\'\i a s/n, E-18008 Granada, Spain \\
$^{5}$NSF's NOIRLab, 950 N.\ Cherry Ave., Tucson, AZ 85719, USA \\
$^{6}$Department of Physics and Astronomy, Western Kentucky University, Bowling Green, KY 42101, USA \\
$^{7}$School of Physics and Astronomy, Sun Yat-Sen University, Zhuhai 519082, China}


\begin{abstract}
We report spectroscopic surveys of planetary nebulae (PNe) in the 
Milky Way and Andromeda (M31), using the 10.4-m Gran Telescopio 
Canarias (GTC).  The spectra are of high quality and cover the whole 
optical range, mostly from 3650\,{\AA} to beyond 1\,$\mu$m, enabling 
detection of nebular emission lines critical for spectral analysis and 
photoionization modeling.  We obtained GTC spectra of 24 compact 
(angular diameter $<$5 arcsec) PNe located in the Galactic 
disk, $\sim$3--20\,kpc from the Galactic centre, and can be used to 
constrain stellar evolution models and derive radial abundance 
gradients of the Milky Way.  We have observed 30 PNe in the outer halo 
of M31 using the GTC.  These halo PNe are uniformly metal-rich and 
probably all evolved from low-mass stars, consistent with the 
conjecture that they formed from the metal-rich gas in M31 disk but displaced to their present locations due to galaxy interactions.
\end{abstract}


\maketitle

\section{Introduction}
\label{sec1}

Planetary nebulae (PNe) evolved from low- to intermediate-mass stars 
($\sim$1--8\,$M_{\odot}$), which account for an absolute majority of 
the stellar populations in the universe.  PNe are ionized shells of 
gas ejected episodically by asymptotic giant branch (AGB) stars during 
the late-stage stellar evolution.  Although belonging to the 
interstellar medium (ISM) and having a very short visible/dynamical 
age ($\sim$10$^{4}$\,yr) due to nebular expansion, PNe are a direct 
link between stellar evolution and the interstellar gas, and can be 
used to well constrain the theory of AGB nucleosynthesis 
\citep[e.g.][]{Fang18,Henry18}. 

Chemical abundances of PNe, both of $\alpha$-elements and of N, C and 
He, can be used to constrain stellar evolutionary models and quantify 
the contribution of low- to intermediate-mass stars to Galactic 
chemical enrichment.  CNO abundances give a direct indication of 
whether the AGB stars have gone through hot bottom burning (HBB) and 
the third dredge-up \citep{Herwig05}, indicating a more massive 
progenitor.  The $\alpha$-elements can also be used to constrain 
Galactic chemical evolution by measuring metallicity gradients in the 
disk \citep[e.g.][]{Milingo10,SH10}.

Abundances of Galactic PNe have been investigated for decades, but mostly with significant limitations.  First, of the current several hundred Galactic PNe with detailed chemical analyses, most are older, extended objects \citep[e.g.][]{Dufour15,Henry18}.  Deriving total elemental abundances in extended PNe using small apertures (e.g.\ narrow slits) involves substantial uncertainties because the ionization is often highly stratified in a PN \citep{OF06}.  It is hard to extrapolate abundances from one part of a PN to the whole nebula.  Compact PNe fit into a long slit, so the flux of the entire nebula is measured, and little correction is needed.  A second limitation on abundance analyses of Galactic PNe is that results between similar studies by different observations/authors are frequently inconsistent, due to different ionization correction factors (ICFs) adopted.  ICFs are used to estimate abundances of ions that are absent in optical spectra \citep{KB94}, and work generally well; but since they were never calibrated for extreme metallicity, electron temperature, or ionization levels due to lack of infrared (IR) or UV data at the time, the use of ICFs introduces larger uncertainties in those regimes. Hence, the best way to address this problem is optical spectroscopy in combination with the IR/UV data.  That is why we carried out optical spectroscopy of Galactic compact PNe with available \textit{Spitzer} mid-IR and \textit{HST}/STIS UV observations (see Section~\ref{sec2}). 

The main-sequence progenitors of PNe encompass a wide range of stellar 
ages, corresponding to a broad range in stellar population.  Given 
their bright, narrow emission lines, PNe are easily detectable in 
distance galaxies/clusters 
\citep[e.g.][]{Gerhard05,Gerhard07,Longobardi18}, 
and are excellent tracers of the chemistry, dynamics and stellar 
population of host galaxies.  The Andromeda Galaxy (M31) is the 
nearest (785~kpc; \citealt{McConnachie05}) large disk system and best 
candidate for studying galaxy merger and evolution.  Numerous 
large-scale substructures (i.e.\ stellar streams; e.g.\ 
\citealt{Ibata01}) as well as inhomogeneity in metallicity have been 
revealed in M31's extended halo by panoramic surveys such as 
PAndAS \citep{McConnachie09}, pointing to a tumultuous merging history 
of this galaxy. 

One long-standing unresolved question is what the origin of M31's 
stellar substructure is.  It has been proposed that the Northern Spur 
and the Southern Giant Stream, two very prominent substructures, might 
be connected by a stellar stream \citep{Ibata01,Merrett03}, but this 
hypothesis needs assessment.  Recent hydrodynamical simulations 
suggests that a single major merger might be responsible for the bulk 
of the substructures in the M31 halo, including the Southern Giant 
Stream \citep{Hammer18}.  However, these simulations, including all 
previous efforts, are still somewhat speculative; accurate 
observations yet to be used help constrain the modeling.  PNe are the 
only ISM/nebulae that exist in almost every part of a galaxy, from the 
disk to the bulge as well as the outer halo; the nebular emission 
lines of a PN can be measured to derive accurate ionic/elemental 
abundances.  PNe thus can be used as a tracer to study the properties 
of M31 halo substructures.  This is the main driving science of our 
GTC spectroscopy of PNe in M31's halo (see Section~\ref{sec3}).

\section{Spectroscopic Survey of compact PNe in the Galactic Disk} 
\label{sec2}

We carried out deep optical spectroscopy of 24 compact (angular 
diameter $<$5~arcsec) PNe in the Galactic disk, using the OSIRIS 
spectrograph on the 10.4\,m Gran Telescopio Canarias (GTC, La Palma). 
The targets are mostly Northern objects, and were carefully selected 
from a well defined sample of 150 compact PNe whose mid-IR 
spectra were obtained with \textit{Spitzer}/IRS 
\citep{Stanghellini12}.  They cover an adequate galactocentric range 
($\sim$3--20\,kpc) so that radial metallicity gradients can be readily 
available. 

\begin{figure}
\begin{center}
 \includegraphics[width=11cm]{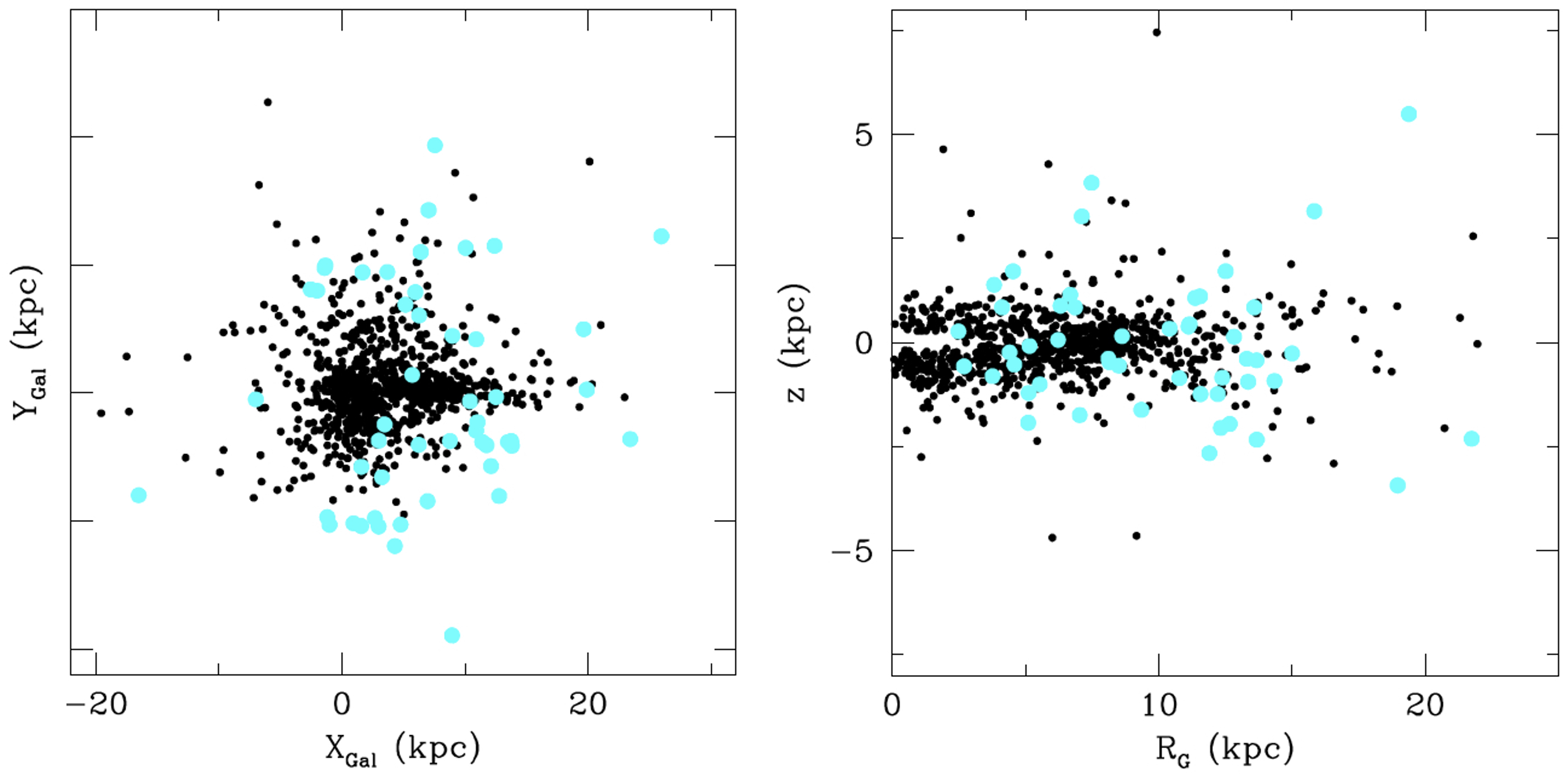}
\end{center}
\caption{Spatial distribution of compact PNe (large, cyan dots) 
against the general distribution of Galactic PNe (small, black dots).  
\textit{Left}: top-view of the Galactic disk.  \textit{Right}: 
galacocentric distance vs. distance from the Galactic plane.  Images 
adopted from \citet[][Figures\,8 and 9 therein]{Stanghellini16}.}
\label{GalacticPNe}
\end{figure}

\begin{figure}
\begin{center}
 \includegraphics[width=13.25cm]{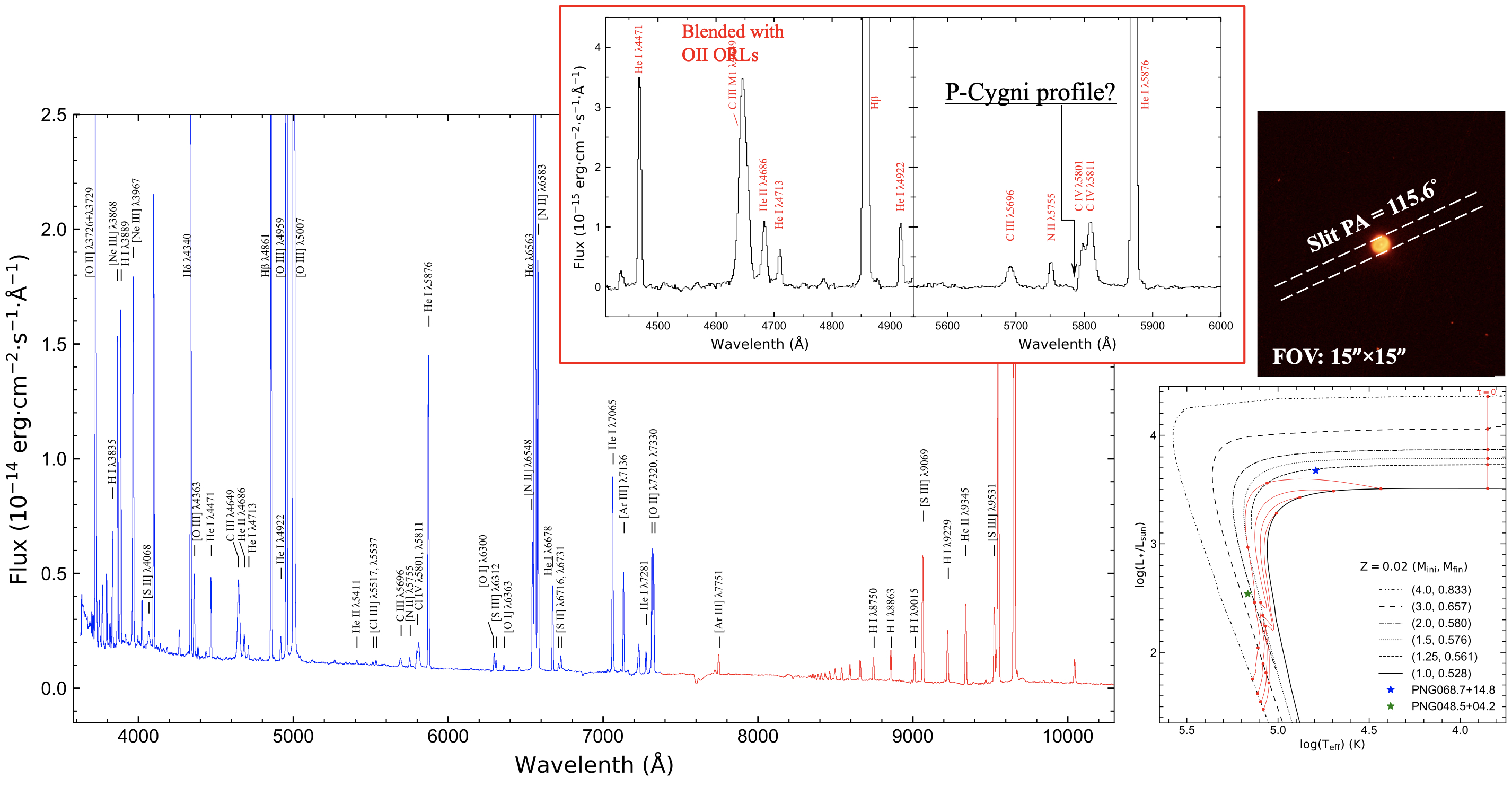}
\end{center}
\caption{\textit{Left}: GTC OSIRIS long-slit spectrum of Galactic 
planetary nebula PN\,G068.7$+$14.8; blue and red represent the R1000B 
and R1000R grisms, respectively.  \textit{Right-top}: $HST$ WFC3 F502N 
narrowband image of PN\,G068.7$+$14.8 in the [O~{\sc iii}] 
$\lambda$5007 emission line, adopted from \citet{Stanghellini16}; the 
GTC long slit (white-dashed lines) with 1-arcsec width is overlaid. 
\textit{Right-bottom}: Central star position of PN\,G068.7$+$14.8 
(blue star) in H-R diagram, where model tracks of the H-burning 
post-AGB sequences calculated by \citet{Bertolami16} at $Z$ = 0.02 are 
overplotted, different line types representing different initial and 
final masses.  The red curves are the isochrones for evolutionary ages 
($\tau$ = 0, 5000, 10,000, 15,000, and 20,000 years) since the 
beginning of post-AGB defined at log$T_{\rm eff}$ = 3.85.  The central 
star position of another Galactic compact nebula PN\,G048.5$+$04.2 
(green star) is also presented for purpose of comparison of a 
different evolutionary status from that of PN\,G068.7$+$14.8. 
\textit{Top-inset}: Zoom-in of the GTC spectrum showing the C~{\sc 
iii} and C~{\sc iv} lines possibly emitted from the central star, with 
a likely P-Cygni profile marked (Huang \& Fang, in preparation).} 
\label{PN_spectrum}
\end{figure}

The GTC/OSIRIS observations was obtained from June to July in 2016 
(GTC program No.: GTC66-16A; PI: X.\ Fang) in the long-slit 
spectroscopy mode, with 1~arcsec slit width.  The blue and red grisms, 
R1000B and R1000R, were used, covering spectral ranges 
$\sim$3630--7850\,{\AA} and $\sim$5080--10370\,{\AA}, respectively. 
The OSIRIS detector consists of two CCDs with 2048$\times$4096 
pixels\footnote{In 2023, OSIRIS was updated to OSIRIS$+$, with a new 
monolithic 4k$\times$4k CCD installed. \url{https://www.gtc.iac.es/instruments/osiris+/osiris+.php}}. 
The size of a single pixel is 15 $\mu$m, corresponding to angular 
scale of 0.127~arcsec; the standard observing mode was used where the 
output images were binned by 2$\times$2.  Spectroscopic observations 
were carried out in the dark moon night with a wonderful seeing of 
0.6--0.8~arcsec.  During observations, multiplet exposures were made 
for each PN for purpose of cosmic-ray removal and increasing the 
signal-to-nose ratio.  In total, $\sim$40\,hr observations were 
completed at GTC for the 24 compact PNe. 

Reduction of the GTC OSIRIS spectra generally follows the standard 
procedure for the long-slit spectra, using {\sc iraf}\footnote{{\sc 
iraf}, the Image Reduction and Analysis Facility, is distributed by 
the National Optical Astronomy Observatory, which is operated by the 
Association of Universities for Research in Astronomy under 
cooperative agreement with the National Science Foundation.}. 
As an example, we present in Figure~\ref{PN_spectrum} the final 
reduced, calibrated and extracted 1D spectrum of one of our targets, 
PN\,G068.7$+$14.8, where C~{\sc iii} $\lambda\lambda$4649,5696 and 
C~{\sc iv} $\lambda\lambda$5801,5811 broad emission lines, probably 
coming from the PN central star, are well detected.  We analyzed the 
nebular spectra, and carried out photoionization modeling using the 
{\sc Cloudy} code \citep{Ferland98,Ferland17} to derived the PN 
central star properties (e.g.\ \textit{bottom-right} panel in 
Figure~\ref{PN_spectrum}). 

Complete analyses of the GTC spectra of all compact PNe targets, in 
combination with the archival \textit{HST}/STIS UV--optical and 
\textit{Spitzer} min-IR data, are underway (Fang et al.\ 2023a, in 
preparation).  The more reliable ICFs of different elements will be 
derived for our PNe using the multi-wavelength spectra.  {\sc Cloudy} 
photoionization models, with an aid of the \textit{HST} optical 
images \citep{Stanghellini16}, will be developed based on the 
UV-optical-IR spectra, to derive the central star masses of our sample 
PNe and consequently, the main-sequence masses/ages.  We will compare 
the chemical abundances with predictions from the stellar evolution 
models -- in particular the AGB nucleosynthesis models 
\citep[e.g.][]{KL16}, and correlate them with the dust chemistry, 
metallicity and central star masses. 

Given the distribution (Galactic $l$ and $b$, distance) of our sample 
in the Galaxy, it is suitable to explore the nebular and stellar 
properties across the Galactic disk.  Our Northern sample is 
complemented by the previous observations of the Southern objects 
(using the 4.1-m SOAR telescope) and other published spectroscopy.  
With the available \emph{Gaia} distances, we will set strong 
constraints on the Galactic evolutionary models through the analysis 
of chemical (oxygen and other $\alpha$-elements) and population 
gradients using these compact PNe.

\section{Spectroscopic Survey of the PNe in M31 halo} 
\label{sec3}

\begin{figure}
\begin{center}
 \includegraphics[width=13.25cm]{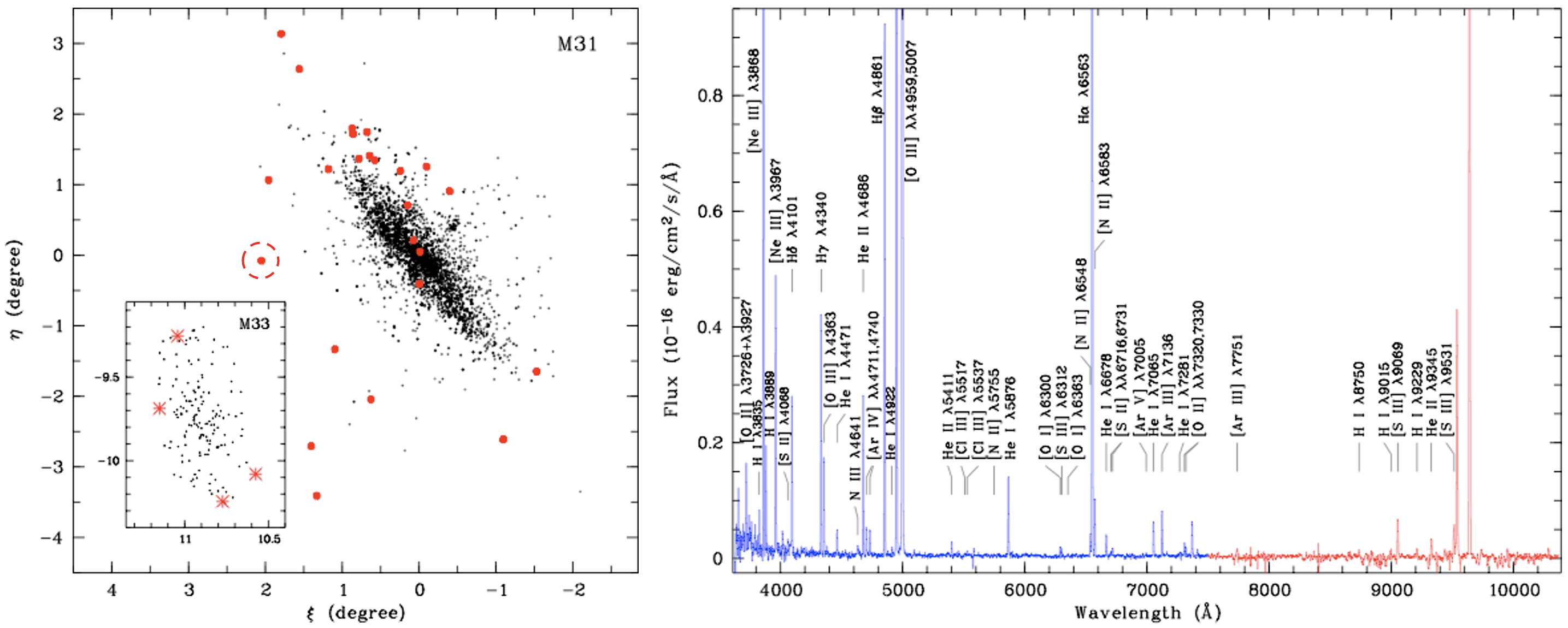}
\end{center}
\caption{\textit{Left}: Distribution of PNe in M31$+$M33.  The PNe 
targeted by our GTC observations \citep[][and several unpublished 
objects]{Fang13,Fang15,Fang18} are highlighted in red-filled circles 
and red asterisks.  The inset shows the PNe in M33 in M31-centred 
coordinates.  \textit{Right}: GTC OSIRIS long-slit spectrum of a 
halo PN in M31 (marked by a red-dashed circle in the \textit{left} 
panel) obtained with the R1000B (blue) and R1000R (red) grisms, with 
emission lines labeled.}
\label{M31_PNe}
\end{figure}

Since 2012, large-aperture (8--10\,m class) optical telescopes have 
been used to obtain the spectra of PNe in M31, mostly in M31's outer 
disk \citep{Kwitter12,Balick13,Corradi15}, and all have oxygen 
abundances close to the solar level. 
Spectroscopic analysis of the outer-halo PNe in M31 were extremely 
scarce.  Our first attempt of optical spectroscopy of bright PNe in 
the outer halo in the Northern Spur substructure) of M31 was carried 
out using the 5.1-m Hale telescope, only with limited data quality, 
although our analysis revealed metal-rich nature of the halo objects 
\citep{Fang13}.  Since 2014, we continued this effort using the 
10.4-m GTC in La Palma. 
So far, over 100~hr observations have been completed at the GTC for 
$\sim$30 PNe in M31's halo (Figure~\ref{M31_PNe}), mostly with 
excellent spectral quality (e.g.\ 
Figure~\ref{M31_PNe}-\textit{right}).  Our halo samples 
kinematically deviate from the disk population, and were carefully 
selected so that they are located in different halo substructures, 
covering a vast spatial extension 
(Figure~\ref{M31_PNe}-\textit{left}).  The instrument setup 
(spectrograph, grisms, slit width and observing mode) and procedure 
of data reduction are the same as introduced in Section~\ref{sec2}. 

\begin{figure}
\begin{center}
 \includegraphics[width=12.5cm]{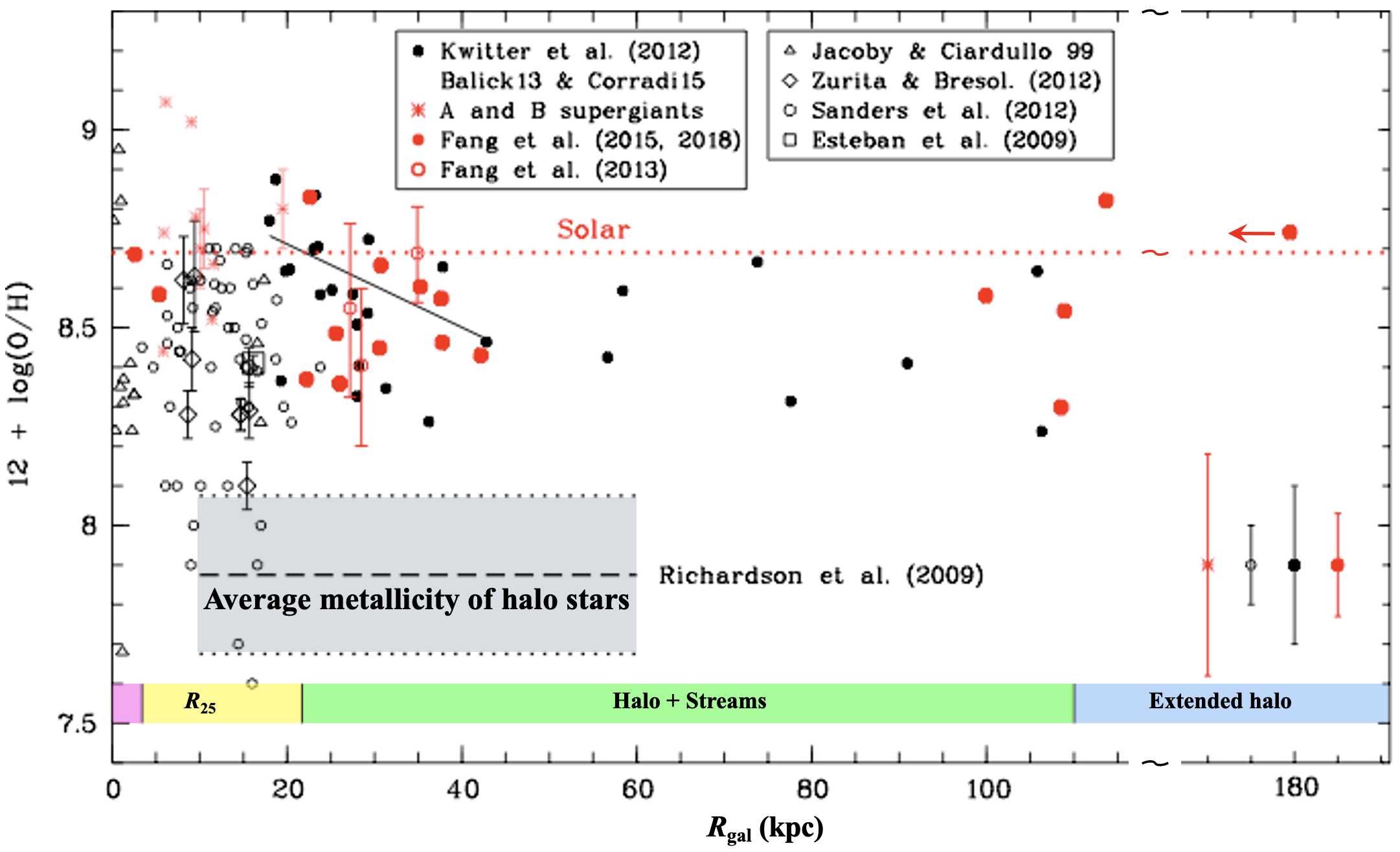}
\end{center}
\caption{Radial distribution of oxygen abundances in M31 (figure 
adopted from \citealt{Fang18}).  The M31 halo PNe previously 
targeted by our GTC observations are red-filled circles (see 
Figure\,\ref{M31_PNe}-\textit{left}).  The M31 outer-disk PNe 
(observed by \citealt{Kwitter12}, \citealt{Balick13}, and 
\citealt{Corradi15}) are black dots.  Other literature samples of 
PNe, H~{\sc ii} regions and supergiants in M31 are over-plotted (see 
legend).  The black-solid straight line is a linear fit to the disk 
PNe between 20 and 40\,kpc \citep{Kwitter12}.  The red-dotted line 
marks the solar value.  The horizontal black-dashed and dotted lines 
represent the mean metallicity and dispersion (also grey-shaded) of 
halo stars between 10 and 60\,kpc \citep{Richardson09}.}
\label{radial_oxygen}
\end{figure}

All the outer-halo PNe of M31 we have published so far also have 
oxygen abundances close to the Sun, [O/H]$\gtrsim\,-$0.4 with 
modest scatter \citep[][see also 
Figure~\ref{radial_oxygen}]{Fang13,Fang15,Fang18}, and probably all 
evolved from low-mass ($<$2\,$M_{\odot}$) stars mostly with ages of 
$\sim$2--5\,Gyr.  Our halo sample cannot be distinguished from the 
disk population in metallicity or the main-sequence mass/age 
\citep{Fang18}, which is in stark contrast with the underlying, 
smooth, metal-poor halo of M31 (Figure~\ref{radial_oxygen}).  We 
conjecture that these halo PNe might have originally formed from 
metal-rich gas in M31 disk but were displaced to the halo regions and 
gained their current stream/substructure kinematics as a result of 
M31's interaction with its satellite(s).  This astrophysical picture, 
although highly speculative, agrees with our present-day knowledge 
that M31's extended halo evolved from complex galactic mergers and 
interactions.  Further investigation using a combined sample in 
M31$+$M33 is underway (Fang et al.\ 2023b, in preparation). 

Photoionization models will be constructed for the halo/substructure 
PNe in M31 to derive their main-sequence masses ($M_{\rm ini}$).  We 
will then derive the He/H and N/O abundance ratios versus $M_{\rm 
ini}$ relations for the combined sample of the halo/substructure PNe 
in M31 together with the Galactic compact PNe observed using GTC (see 
Section~\ref{sec2}).  These abundance-mass relations will be used to 
constrain the AGB models.  The surface contents of helium and nitrogen 
are expected to increase as a consequence of the second dredge up and 
HBB that occur in intermediate-mass AGB stars 
with $M_{\rm ini}\,\gtrsim$3--5\,$M_{\odot}$\citep[e.g.][]{KL16}. 
However, previous spectroscopic analysis of M31 and Galactic PNe 
indicates that HBB might actually occur at $<$3\,$M_{\odot}$.  Our new 
observations will be used to carefully assess this (Fang et al.\ 
2023b, in preparation).


\end{document}